\documentclass{emulateapj}
\usepackage{apjfonts}
\slugcomment{Accepted for publication in ApJ}
\shorttitle{Relativistic neutrons for AGN jet production}
\shortauthors{Toma \& Takahara}
\begin{document}
\title{
Baryon Loading of AGN Jets Mediated by Neutrons
}
\author{K. Toma, and F. Takahara}
\affil{Department of Earth and Space Science, Graduate School of Science, Osaka University,
Toyonaka 560-0043, Japan \\ toma@vega.ess.sci.osaka-u.ac.jp}
\begin{abstract}
Plasmas of geometrically thick, black hole (BH) accretion flows in active 
galactic nuclei (AGNs) are generally collisionless for protons, and involve magnetic field 
turbulence. Under such conditions a fraction of protons can be accelerated stochastically 
and create relativistic neutrons via nuclear collisions. These neutrons can freely escape 
from the accretion flow and decay into protons in dilute polar region above the 
rotating BH to form relativistic jets. 
We calculate geometric efficiencies of the neutron energy and mass 
injections into the polar region, and show that this process can deposit luminosity as high
as $L_j \sim 2 \times 10^{-3}\;\dot{M} c^2$ and mass loading $\dot{M}_j \sim 6 \times 10^{-4}\;\dot{M}$
for the case of the BH mass $M \sim 10^8\;M_{\odot}$, where $\dot{M}$ is mass 
accretion rate. The terminal Lorentz factors of the jets are $\Gamma \sim 3$, and
they may explain the AGN jets having low luminosities. For higher luminosity jets, 
which can be produced by additional energy inputs such as Poynting flux, 
the neutron decay still can be a dominant 
mass loading process, leading to e.g., $\Gamma \sim 50$ for $L_{j,{\rm tot}} \sim 3 \times 10^{-2}\;\dot{M}c^2$.
\end{abstract}


\section{Introduction}
\label{sec:intro}

One of the major problems in astrophysics is the production mechanism of relativistic
jets. They are associated with active galactic nuclei (AGNs), Galactic black hole (BH)
candidates, and gamma-ray bursts (GRBs). It is inferred that AGN jets have Lorentz 
factors of $\Gamma \sim 10-100$, luminosities as high as 
the Eddington luminosity $L_{\rm Edd}$, and opening angles of $\theta_j \sim \Gamma^{-1}$.
Although the matter content of AGN jets is still an open problem, the inertia is seemingly 
dominated by protons \citep{sikora05}. The masses of central BHs of AGNs are typically 
$M \sim 10^7~M_{\odot} - 10^9~M_{\odot}$.

An outflow can have a relativistic velocity if the enthalpy per unit rest energy
$\mu \equiv L_{j,{\rm tot}}/\dot{M}_j c^2 \gg 1$ in the vicinity of the BH, where 
$L_{j,{\rm tot}} = L_B + L_k$ is the total luminosity, $L_B$ and $L_k$ are Poynting and 
particle kinetic luminosities, respectively, and $\dot{M}_j$ is mass loading rate. 
If the source of $L_{j,{\rm tot}}$ is gravitational energy of the accreting mass on the 
central BH, we have $L_{j,{\rm tot}} < GM\dot{M}/R < \dot{M} c^2$, where $\dot{M}$ is
mass accretion rate, and radius $R$ should be larger than $GM/c^2$.
This leads to $\dot{M}/\dot{M}_j \gg 1$, implying that the production mechanism of
relativistic jets should be converting gravitational energy into Poynting and/or 
kinetic energies, and concentrating them on a small fraction of mass.

It is therefore likely that the Poynting and/or kinetic energies are injected in
polar region above the rotating BH, `the funnel', where mass loading is exponentially 
suppressed by the centrifugal barrier \citep{abramowicz78}. 
This configuration is also suitable for the outflow to be collimated by the external pressure. 
The magnetically dominated jet models, or magnetohydrodynamic (MHD) models, have been
recently progressed with numerical simulations \citep[e.g.,][]{mckinney06}. 
If a certain amount of particles are injected (or generated) in the funnel, 
the strong poloidal magnetic fields associated with electric currents flowing in the 
accretion torus accelerate the flow of the particles. The MHD flow can accelerate
to a relativistic velocity if suitable boundary conditions are satisfied
\citep{komissarov07,lyubarsky09,granot11}. On the other hand, the kinetically dominated jet 
models, so-called fireball models, have been actively discussed for GRB jets, for 
which the energy injection via $\nu \bar{\nu}$ annihilation could be efficient 
\citep{zalamea11}. A thermally dominated spherical blob of gas inevitably accelerates to a relativistic 
velocity \citep{meszaros93,piran93,kobayashi99}. It is even possible that large amount 
of remaining thermal energy at the photosphere is released as prompt $\gamma$-ray 
emission itself \citep[e.g.,][and references therein]{toma11}.  For AGN jets also, fireball models
have been elaborately studied by some authors \citep[e.g.,][]{asano07,asano09,becker11}.

However, the above models do not answer a question ``Why are the Lorentz factors
of AGN jets regulated to $\Gamma \sim 10-100$? Why not $\Gamma \gg 100$ or
$\Gamma \sim$ a few?''
In either of the MHD or fireball model, mass loading rate into the funnel is essential for 
determining the final Lorentz factor (and the radiation properties) of jets. 
There is an interesting idea on this point for GRB jets. Their mass loading might be 
determined by neutron diffusion from the disk wind surrounding the jet 
\citep{levinson03,mckinney05}. The jet is considered to have globally ordered magnetic
fields that may suppress proton diffusion across the jet-disk wind 
boundary.\footnote{See also \citet{levinson06} and \citet{ioka10} for other ideas on
relativistic jet mass loading.}

In this paper, we discuss the role of the neutrons for mass (as well as energy) injection
of AGN jets. We focus on relativistic neutrons escaping from the accretion flow, and
calculate the fraction of the total escaping neutrons that decay into protons 
in the funnel. The processes of the relativistic neutron escape 
from the BH accretion flow have been discussed by several authors
\citep[e.g.,][]{eichler78,sikora89,begelman90,atoyan92,contopoulos95},
although the efficiencies of mass and energy injections in the funnel region 
have not been studied in detail. In Section~\ref{sec:scenario}, we make a short review on 
proton acceleration and cooling in the accretion flow, and parametrize the spectrum of the escaping 
neutrons. Then we formulate the geometric injection efficiency in Section~\ref{sec:formulation}, 
and show the results of $L_j/\dot{M} c^2$ and $\dot{M}_j/\dot{M}$ 
in our model in Section~\ref{sec:results}. 
Conclusion and discussion are given in Section~\ref{sec:discussion}.

\section{Relativistic Neutron Production and Escape}
\label{sec:scenario}

We consider the AGN central engine as follows. A rotating BH has an accretion flow around it.
The accretion flow is geometrically thick, and creates a hot corona and/or non-relativistic wind 
extending vertically, although the polar region above the central BH is dilute because of the 
centrifugal barrier. There can be large-scale magnetic fields anchored on the accretion
flow, which may also suppress protons leaking into the funnel, since the Larmor radius
\begin{equation}
R_{\rm L} = \frac{\gamma_p m_p c^2}{eB} \simeq 3.1 \times 10^3\;\gamma_p B_3^{-1}\;{\rm cm}
\end{equation}
is much smaller than the system size characterized by the Schwarzschild radius
\begin{equation}
R_{\rm s} = \frac{2GM}{c^2} \simeq 3.0 \times 10^{13}\;M_8\;{\rm cm},
\end{equation}
where $B_3 = B/(10^3\;{\rm G})$ and $M_8 = M/(10^8\;M_{\odot})$. Neutrons are 
not affected by the magnetic fields, so that they can be an effective source of mass injection into
the funnel.

One may consider the origin of such neutrons as thermal process in the accretion
flow. The temperature of protons and heavy nuclei is as high as 
$k T_p \sim G M m_p/R \sim 500 R_*^{-1}\;$MeV, where $R_* \equiv R/R_{\rm s}$. 
Then the helium breakup and the $pp$ collisions can produce thermal neutron 
component. The neutron fraction can be $n_n/(n_n+n_p) \lesssim 0.1$ in the vicinity of the 
BH \citep[e.g.,][]{filho03,hu08}. However, non-relativistic neutrons only survive as long as
\begin{equation}
D \equiv c \tau_n \sim 3 \times 10^{13}\;{\rm cm},
\end{equation}
where $\tau_n \sim 10^3\;$s is the decay time of a neutron. This is comparable to the system
size characterized by $R_{\rm s}$, and thus they cannot reach the funnel in the case of 
$M \gtrsim 10^8\;M_{\odot}$.

We then focus on relativistic neutrons created in the accretion flow. 
The plasmas of the geometrically thick AGN accretion flows are generally collisionless for
protons \citep[cf.][]{takahara85}. In such plasmas, the proton energy distribution is 
not necessarily Maxwellian, and some fraction of the protons can be accelerated 
to relativistic speeds. The relativistic hadrons produce relativistic neutron 
component via the $pp$ and/or $p\gamma$ collisions \citep{sikora89,begelman90}. 
The relativistic neutrons with Lorentz factor $\gamma_n$ survive over
$D \gamma_n \simeq 3 \times 10^{13}\;\gamma_n\;$cm, so that they can reach
the funnel and decay into relativistic protons (and electrons) there. The magnetic
coupling in the funnel will isotropize them, leading to the electron-proton fireballs.
Neutrons that decay outside the funnel just energize the dense non-relativistic wind.

As for the proton acceleration mechanism in the geometrically thick accretion flows, 
there are many possibilities, and we do not specify it in this paper.
 If the accretion results from turbulent viscosity driven by magneto-rotational
instability \citep{balbus91}, the plasma involves magnetic field fluctuations 
and reconnections, where protons can be accelerated via second-order
and/or first-order Fermi processes \citep[e.g.,][]{dermer96,hoshino12,riquelme12}. 
The flows may also have regions with strong (less-fluctuated) magnetic fields, where 
accretion can result from angular momentum transport by magnetically driven winds.
Some fields may be oppositely directed, giving rise to magnetic reconnections, which
can accelerate particles via first-order Fermi process \citep[e.g.,][]{drury12,degouveia10,vieyro12}. 
Furthermore, protons could be accelerated at a standing shock formed in the 
accretion flow \citep{becker11}.

In the following, we make a short review of \citet{sikora89} and \citet{begelman90}
for parametrizing the neutronization factor and the escaping neutron spectrum.

\subsection{Spectrum of escaping neutrons}
\label{subsec:spectrum}

Relevant processes for protons are the accretion inflow, Fermi acceleration, 
$p \gamma$ cooling, and $pp$ cooling, and those for created neutrons are the escape
from the inflow, $n \gamma$ cooling, and $np$ cooling. The proton escape or spatial 
diffusion is negligible because of the small Larmor radius, whereas protons are advected 
on the inflow timescale $t_{\rm in}$:
\begin{equation}
t_{\rm in} \sim \frac{R}{v_{\rm in}} = 10\;v_{{\rm in},-1}^{-1} \frac{R}{c},
\end{equation}
where $v_{\rm in} = 0.1 v_{{\rm in},-1} c$ is the accretion fluid velocity.
Fermi acceleration timescale is estimated as 
$t_{\rm acc} \sim \xi R_L/c \simeq 1.0 \times 10^{-3}\;\gamma_p \xi_4 B_3^{-1}\;$s, 
where $\xi = 10^4 \xi_4$ is the factor related to the magnetic field fluctuations. 
This indicates that the acceleration is so efficient that protons can be highly
energized. At high energy ranges, the $p \gamma$ collisions are the most important 
cooling process, where we assume a typical radiation field with luminosity 
$10^{-4} L_{\rm Edd} < L_r < L_{\rm Edd}$ and spectrum $F_{\nu} \propto \nu^{-\alpha}$ with 
$0.5 < \alpha < 1.5$. 
This assumed radiation field is rather generic, which may include thermal 
as well as non-thermal emission of electrons.
The equality of $t_{\rm acc}$ with the $p \gamma$ 
cooling timescale $t_{p\gamma}$ leads to the maximum proton Lorentz factor
$\gamma_{p,M} \sim 10^7 (B_3 R_* /\xi_4)^{1/2} (L_r/10^{-2} L_{\rm Edd})^{-1/2}$.

For the BH mass $M \sim 10^7\;M_{\odot} - 10^9\;M_{\odot}$, the system size is
$\lesssim 10^{15}\;$cm, so that the protons with $\gamma_p \lesssim 10^2$ 
(which are converted into neutrons with similar Lorenz factor $\gamma_n$) are important. 
At such energy ranges, the $pp$ collisions are more efficient for creation of 
neutrons than the $p\gamma$ collisions. The $pp$ cooling time for protons is 
estimated as
\begin{equation}
t_{pp} \sim \frac{1}{n_p c \sigma_{pp} K_{pp}} \sim 30\;\tau_p^{-1} \frac{R}{c},
\label{eq:t_pp}
\end{equation}
where $n_p$ and $\tau_p = \sigma_T n_p R$ are the proton number density and 
the Thomson optical depth at radius $R$, respectively. The $pp$ cross section is 
$\sigma_{pp} \sim \sigma_T/17$, and $K_{pp} \simeq 1/2$ is the inelasticity.
If $t_{pp} < t_{\rm in}$, the $pp$ collisions occur efficiently. 
For $t_{\rm in} < t_{pp}$, the efficiency reduces by a factor of $t_{\rm in}/t_{pp}$.

The created neutrons are not magnetically coupled to the background plasma.
They escape without being absorbed if  $t_{\rm esc}^{(n)} \sim R/c$ is smaller
than the timescales of $np$ and $n\gamma$ collisions, $t_{np}$ and $t_{n\gamma}$,
that are similar to $t_{pp}$ and $t_{p\gamma}$, respectively. The maximum Lorentz
factor of the escaping neutrons is given by $t^{(n)}_{\rm esc} = t_{n\gamma}$, as
$\gamma_{n,M} \sim 10^6 (L_r/10^{-2} L_{\rm Edd})^{-1}$.

At lower energy ranges, say $\gamma_p < 10^5$, we have 
$t_{\rm acc} < t_{\rm esc}^{(n)} < {\rm min}(t_{\rm in}, t_{pp}) < {\rm max}(t_{\rm in}, t_{pp}) 
< t_{p\gamma}$, for $\tau_p < 20$. In this case, neutrons created by $pp$ collisions 
freely escape from the accretion flow. 
The kinetic equations for the proton and neutron number densities in the steady
state and for a power-law proton injection function $\dot{N}_p \propto \gamma_p^{-p}$
indicate that the ratio of the escaping neutron number flux $\dot{N}_n$ to the proton 
injection flux is estimated as
\begin{equation}
f_n \equiv \frac{\dot{N}_n}{\dot{N}_p} \sim \frac{1}{2(p-1)+1} 
{\rm min} \left(1, \frac{t_{\rm in}}{t_{pp}}\right).
\label{eq:f_n}
\end{equation}
The numerical factor $2$ in the denominator is determined by the probability of the 
charge exchange during a single $pp$ collision and $K_{pp}$.
The neutron spectrum is given by $\dot{N}_n \propto \gamma_n^{-p}$ for the 
low-energy range satisfying $t_{pp} < t_{p\gamma}$. It deviates from a single
power-law at high energy, say $\gamma_n > 10^5$ \citep[see Fig.~7 of][]{begelman90}, 
although the neutrons at such high energy ranges are not relevant for our purpose in this paper.

The process of the creation and escape of neutrons with $\gamma_n < 10^5$
is most effective for $t_{\rm esc}^{(n)} < t_{np}$ and $t_{pp} < t_{\rm in}$, i.e.,
\begin{equation}
3 v_{{\rm in},-1} < \tau_p < 20,
\end{equation}
for which we have $f_n \sim 1/[2(p-1)+1]$.
It may be possible that geometrically thick accretion flows satisfy this condition of
$\tau_p$. To confirm it in detail, however, modeling of accretion flows with 
significant neutron energy release is required, which we leave as separate work.

\section{Calculation of the Geometric Efficiency}
\label{sec:formulation}

The luminosity of the escaping neutrons is parametrized by 
$L_n = f_n f_a f_{\rm th} \dot{M} c^2$, where $f_{\rm th}$ is the ratio of the heating rate
of protons to $\dot{M} c^2$, $f_a$ is the ratio of the rate for accelerated protons to 
the heating rate of protons, and $f_n$ is the neutronization factor estimated by Eq.~(\ref{eq:f_n}).
The heating rate can be as high as $\sim (1/2)GM\dot{M}/R_{\rm s} \sim \dot{M}c^2/4$,
i.e., $f_{\rm th} \lesssim 0.3$, while we have
$f_n \lesssim 1/[2(p-1)+1] \lesssim 0.3$ for a reasonable range $p>2$.
It is difficult to estimate the acceleration fraction $f_a$. We only have a rough constraint
$f_a \lesssim 0.3$, which means that the energy density of the accelerated protons will
not dominate that of thermal protons. Therefore we may summarize the microphysical 
efficiency as
\begin{equation}
f_n f_a f_{\rm th} \lesssim 3 \times 10^{-2}.
\end{equation}

The neutrons with this luminosity are released isotropically, and
the protons created via the neutron decays are magnetically coupled and energize
the background plasma. Here we calculate the geometric efficiency, i.e., the fraction 
of neutrons that decay into protons at the polar region. In order to obtain the order 
of magnitude of the geometric efficiency, we set a simple configuration of the BH accretion 
system, and calculations are performed by assuming the Euclidean space.

\begin{figure}
\epsscale{1.1}
\plotone{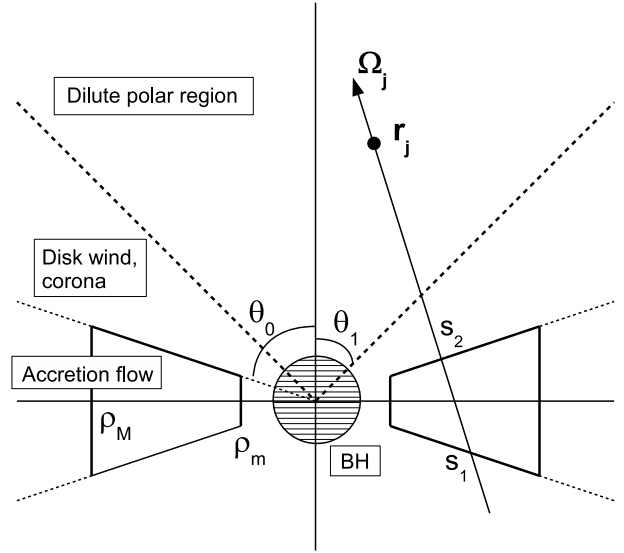}
\caption{
Structure of the BH accretion system for the calculation of the geometric efficiency.
The BH is at the center of the coordinate system. The accretion flow with thickness
$H/R = 1/\tan\theta_0$ has the neutron emission region between the cylindrical radii
$\rho_m = R_{\rm s}$ and $\rho_M = 3R_{\rm s}$. A trajectory of 
neutrons incident to the point ${\bf r}_j = (r_j, \theta_j, \varphi_j)$ is shown.
The dilute polar region is assumed to be a cone with the opening angle $\theta_1$ 
for $r_j < r_2 = 10 R_{\rm s}$, and strongly collimated for $r_j > r_2$.
}
\label{fig:config}
\end{figure}

As illustrated in Figure~\ref{fig:config}, we consider an accretion torus with 
$H/R = 1/\tan\theta_0$ around a BH, which is located at the center of the coordinate 
system. The neutron emission region is assumed as the inner fraction of this torus 
between the cylindrical radii $\rho_m = R_{\rm s}$ and $\rho_M = 3 R_{\rm s}$. 
We assume that the neutron emissivity is uniform in this region for simplicity. 
The spatial volume of this region is $V = (4/3) \pi (\rho_M^3 - \rho_m^3)/\tan\theta_0$. 
Then the neutron emissivity is given by $dE^{(n)}/dt dV d\Omega d\gamma
\equiv j_\gamma^{(n)} = A \gamma^{-p+1}$ for $1 \leq \gamma \leq \gamma_M$,
where $A = L_n/(4\pi V \int^{\gamma_M}_{1} \gamma^{-p+1} d\gamma)$.
Hereafter we will not distinguish the Lorentz factors of protons and neutrons,
since the neutrons with $\gamma_n$ decay into protons with $\gamma_p \sim \gamma_n$.
The dilute polar region is set as the cone with the opening angle $\theta_1$. We assume
that $\theta_1 \sim {\it O}(1)$ for $r_1 = R_{\rm s} < r_j <r_2$, and consider that $r_2$
is a characteristic radius above which the collimation becomes strong due to the pressure 
from the hot corona and/or disk wind. The collimation suppresses the energy injection
for $r_j > r_2$. We will take $r_2 = 10 R_{\rm s}$ as a fiducial case below.

Let us consider a straight line with a parameter $s$ crossing the surfaces of the 
emission region at $s = s_1$ and $s=s_2$. The fractional neutron intensity emitted from the
line element $\Delta \bar{s}$ at $s_1 < \bar{s} < s_2$ and measured at a point $s$ in the
polar region is $\Delta I_\gamma^{(n)} (s) = j_\gamma^{(n)} (\bar{s}) \Delta \bar{s}$, if 
neutron decay is not taken into account. We may approximate the proton intensity 
created at $s_j < s < s_j + ds_j$ through the neutron decay as
$\partial \Delta I_{\gamma}^{(p)}/\partial s_j = \Delta I_{\gamma}^{(n)} (s)
\delta (s_j - \bar{s} - D\gamma)$. Integrating this over $s_1 < \bar{s} < s_2$, we obtain
the proton emissivity created at $s_j$ from the neutrons propagating through a single line as
\begin{eqnarray}
j_\gamma^{(p)} (s_j) \equiv
\frac{\partial I_\gamma^{(p)}}{\partial s_j} (s_j)  
&=& \int^{s_2}_{s_1} j_\gamma^{(n)} (\bar{s}) \delta (s_j - \bar{s} - D\gamma) d\bar{s} 
\nonumber \\
&=& \cases{ 
A\gamma^{-p+1} ~~~(s_j - s_2 < D\gamma < s_j - s_1), \cr
0 ~~~({\rm otherwise})}.
\label{eq:delta}
\end{eqnarray}
For a given point ${\bf r} = {\bf r}_j$ (corresponding to $s_j$), $s_1$ and $s_2$ are 
functions of the incident direction, i.e., $s_1 = s_1(\Omega_j)$ and $s_2 = s_2(\Omega_j)$. 
We obtain the energy injection rate of protons per unit volume at a point ${\bf r} = {\bf r}_j$ 
by integrating $j_\gamma^{(p)}(s_j)$ over solid angle, $dE^{(p)}/dt dV_j d\gamma \equiv 
\dot{u}_\gamma^{(p)}({\bf r}_j) = \int d\Omega_j j_\gamma^{(p)}(s_j)$. This leads to
the mass injection rate per unit volume as 
$dM^{(p)}/dt dV_j d\gamma = \dot{u}_\gamma^{(p)}({\bf r}_j)/(\gamma c^2)$.
Finally we obtain the total energy and mass injection rates by integrating $\dot{u}_\gamma^{(p)}$ 
and $\dot{u}_\gamma^{(p)}/(\gamma c^2)$, respectively, over the total energy range and the 
volume of the polar region. The efficiencies of the neutron energy and mass injections are thus 
written by
\begin{eqnarray}
\left\{ \begin{array}{l}
L_j/(\dot{M} c^2) \\
M_j/\dot{M}
\end{array} \right\}
=
\frac{2\pi}{\dot{M}c^2} 
&&
\int^{\theta_1}_0 d\theta_j \;\sin\theta_j \int^{r_2}_{R_s} dr_j \;r_j^2
\times \nonumber \\
&&
\int^{\gamma_M}_{1} d\gamma \int d\Omega_j
\left\{ \begin{array}{l}
j_\gamma^{(p)} \\
j_\gamma^{(p)}/\gamma
\end{array} \right\},
\end{eqnarray}
where the axisymmetry of this system has allowed us to perform the integration over
the azimuthal angle $\varphi_j$, and we set the upper bound of the $r_j$ integration as
$r_2$ since the energy injection is assumed to be much less effective for $r_j > r_2$
due to strong collimation. The free parameters for calculating the efficiencies
for a given BH mass $M$ are $\tan\theta_0, p, \gamma_M, r_2, \tan\theta_1$, and
$f_n f_a f_{\rm th}$.

\section{Results}
\label{sec:results}

\subsection{Case of $M=10^8\;M_{\odot}$}
\label{subsec:m8}

\begin{figure}
\epsscale{1.1}
\plotone{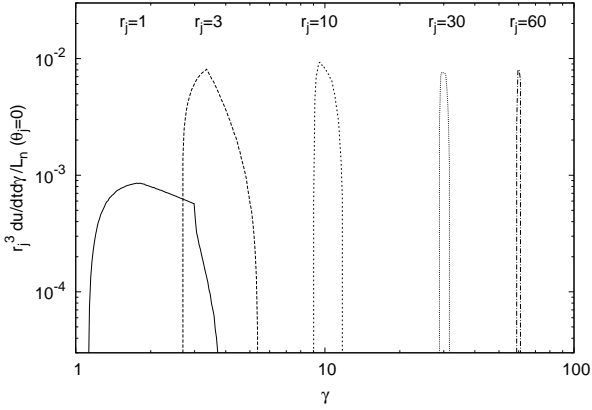}
\caption{
Fraction of spectral energy injection rate per unit volume $\dot{u}_\gamma^{(p)}({\bf r}_j)$ 
(times $r_j^3$) of protons
at the pole $(\theta_j=0)$ of various radii $r_j$, with respect to the escaping neutron 
luminosity $L_n = f_n f_a f_{\rm th} \dot{M} c^2$, for the case of $M_8 = 1$. 
The lines from left to right correspond to $r_j = 1, 3, 10, 30,$ and $60$, respectively 
(measured in unit of $D$). The other parameters are $\tan\theta_0=2, p=2,$
and $\gamma_M=10^2$.
}
\label{fig:ugrt}
\end{figure}

Here we show the calculation results for the case of $M_8 = 1$. In this section we 
measure lengths in unit of $D$. Since $R_{\rm s}$ happens to be equal to $D$
in this case, we have $\rho_m = 1$ and $\rho_M=3$. 
The other parameters are set to be $\tan\theta_0=2, p=2, \gamma_M=10^2, r_2=10,$
and $\tan\theta_1=1$. We will examine the dependence of results on these parameters later.
First, we show the integration result of $\dot{u}_\gamma^{(p)}({\bf r}_j) = \int d\Omega_j j_\gamma^{(p)}(s_j)$
to see the spectral property of the injection rate at various points. We plot $r_j^3 \dot{u}_\gamma^{(p)}/L_n$
at the pole $\theta_j=0$ of various radii $r_j = 1, 3, 10, 30,$ and $60$ in Figure~\ref{fig:ugrt}. 
We can see that the injected protons are distributed over 
relatively broad energy range at $r_j \lesssim 3$, while concentrated 
to $\gamma \sim r_j$ at large radii $r_j \gtrsim 3$. This indicates that for large $r_j$, 
neutrons decay after traversing large distances, so that the detailed structure of 
the emission region is not relevant.

\begin{figure}
\epsscale{1.1}
\plotone{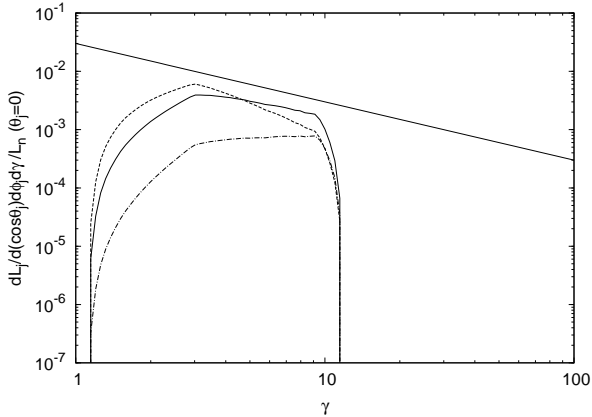}
\caption{
Fraction of spectral energy injection rate per unit solid angle for $1<r_j<r_2=10$
$dL_j/d(\cos\theta_j) d\varphi_j d\gamma = \int^{r_2}_{1} dr_j\; r_j^2 \dot{u}_\gamma^{(p)}$ 
at the pole $(\theta_j=0)$, 
with respect to $L_n = f_n f_a f_{\rm th} \dot{M}c^2$, for the case of $M_8 = 1$
with $\tan\theta_0=2$ and $\gamma_M=10^2$.
The solid, dashed, and dot-dashed lines correspond to the cases of $p=2$, $p=3$, and
$p=1$, respectively. For reference, we plot a power-law function $\propto \gamma^{-1}$
by the solid straight line.
}
\label{fig:ugt}
\end{figure}

Next we show the integration result of $\int^{r_2}_{1} dr_j\; r_j^2 \dot{u}_\gamma^{(p)}$ at
the pole, which is equivalent to  $dL_j/d(\cos\theta_j) d\varphi_j d\gamma$.
We plot this value, normalized by $L_n$, in Figure~\ref{fig:ugt}.
We also show the results for $p=1$ and $p=3$.
For $p=2$, comparing the result with the injection power-law profile $\propto \gamma^{-p+1}$,
we find that the energy is injected efficiently for
$\gamma \sim 3-10$, and has the similar spectrum as the injected one. This property
is explained simply as follows. At the large radii, say $r_j > \tilde{r}_1$, where the detailed 
structure of the emission region is not relevant, the solid angles of the incident neutron 
directions are limited within a small range $\Delta \Omega_j \sim S/r_j^2$, where $S$ is 
the horizontal cross section of the emission region. Also we may write 
$j_\gamma^{(p)}(r_j) = j_\gamma^{(n)} \delta (r_j - \gamma) h$, where 
$h$ represents the mean width of the emission region. Note that we measure
$r_j$ in unit of $D$. These approximations
lead to
\begin{eqnarray}
\frac{1}{L_n} \int^{r_2}_{\tilde{r}_1} dr_j\; r_j^2 \dot{u}_\gamma^{(p)}
&\sim& \frac{1}{4\pi V \int^{\gamma_M}_1 \gamma^{-p+1} d\gamma}
\times \nonumber \\
&&
\int^{r_2}_{\tilde{r}_1} dr_j\; r_j^2 \int d\Omega_j\; \gamma^{-p+1} \delta (r_j - \gamma) h 
\nonumber \\
&\sim& \cases{
\frac{1}{4\pi \int^{\gamma_M}_1 \gamma^{-p+1} d\gamma} \gamma^{-p+1}  ~~~(\tilde{r}_1<\gamma<r_2), \cr
0 ~~~({\rm otherwise}),
}
\label{eq:approx}
\end{eqnarray}
where we set $Sh \sim V$.
Since we have $1/(4\pi \int^{\gamma_M}_1 \gamma^{-p+1} d\gamma) = 0.017$ for $p=2$ and $\gamma_M=10^2$,
this rough calculation well agrees with the 
numerical integration for $3 \lesssim \gamma \lesssim 10$.
For $\gamma < 3$, the spatial volume of emitting points that can connect to the 
pole with decay lengths of $\sim \gamma$ is smaller, so that the injected energy is smaller.

\begin{figure}
\epsscale{1.1}
\plotone{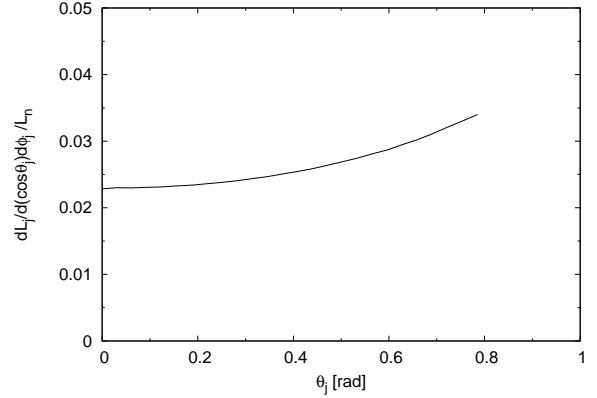}
\caption{
Fraction of energy injection rate per unit solid angle for $1<r_j<r_2=10$
($dL_j/d(\cos\theta_j) d\varphi_j = \int^{\gamma_M}_1 d\gamma
\int^{r_2}_{1} dr_j\; r_j^2 \dot{u}_\gamma^{(p)} $)
with respect to $L_n = f_n f_a f_{\rm th} \dot{M}c^2$ 
as a function of $\theta_j$ for the case of $M_8 = 1$ with the other parameters
$\tan\theta_0 = 2, p=2,$ and $\gamma_M=10^2$.
}
\label{fig:ut20}
\end{figure}

\begin{figure}
\epsscale{1.1}
\plotone{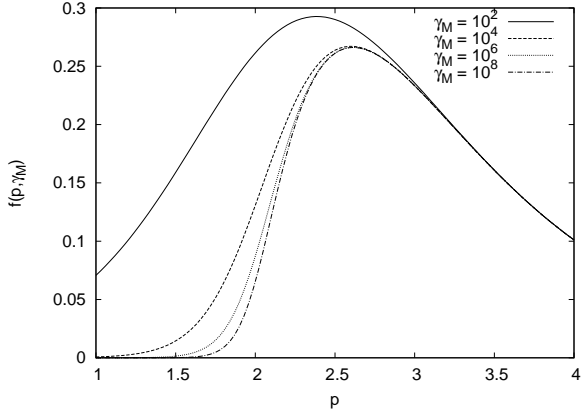}
\caption{
Factor $\int^{10}_{3} \gamma^{-p+1} d\gamma/\int^{\gamma_M}_1 \gamma^{-p+1} d\gamma$
is plotted as a function of $p$ for various $\gamma_M$.
}
\label{fig:eff}
\end{figure}

We integrate $\frac{1}{L_n} \int^{r_2}_{1} dr_j\; r_j^2 \dot{u}_\gamma^{(p)}$ over 
$1<\gamma<\gamma_M = 10^2$ and then obtain $\frac{1}{L_n} dL_j/d(\cos\theta_j)d\varphi_j = 0.023$ at $\theta_j=0$. 
We also show the calculation results of  $\frac{1}{L_n} dL_j/d(\cos\theta_j)d\varphi_j$ as a function 
of $\theta_j$ in Figure~\ref{fig:ut20}.
This shows that the energy injection rate does not 
strongly depend on $\theta_j$. The slight increase for larger $\theta_j$ is due to 
larger contribution for the energy range $\gamma < 3$, but this contribution is 
confirmed to be not significant for the total energy injection $L_j$ in the case of
$\tan\theta_1 = 1$.

Finally we integrate $\frac{1}{L_n} dL_j/d(\cos\theta_j)d\varphi_j$ over $0<\theta_j<\theta_1=\pi/4$ and
multiplied by $2\pi$, obtaining $L_j/L_n = 0.051$. This provides the
total energy injection efficiency for the parameter values 
$\tan\theta_0=2, p=2, \gamma_M=10^2, r_2=10, \tan\theta_1=1$ as
$L_j/(\dot{M}c^2) = 0.051\;f_n f_a f_{\rm th}$.

The geometric efficiency is roughly estimated by integrating Eq.~(\ref{eq:approx})
with $d(\cos\theta_j) d\varphi_j d\gamma$,
\begin{equation}
\frac{L_j}{L_n} \sim 
\frac{2\pi (1-\cos\theta_1)}{4\pi} \frac{\int^{r_2}_{\tilde{r}_1} \gamma^{-p+1} d\gamma}
{\int^{\gamma_M}_1 \gamma^{-p+1} d\gamma},
\label{eq:formula}
\end{equation}
which gives $\simeq 0.04$, if $\tilde{r}_1=3$ is taken. The slight difference
between this rough estimate and the numerical calculation comes from
treating the energy injection rate as constant over $\theta_j$ and neglecting
the contribution for $\gamma < 3$. The above rough estimate is useful, which simply
consists of the two factors, the ratio of  the solid angle of the polar region to $4\pi$ and
the ratio of the energy in the effective injection range $\tilde{r}_1 < \gamma < r_2$ to
the total range $1< \gamma < \gamma_M$.

We confirm that the difference of the torus thickness $\tan\theta_0$ does not substantially
change the value of the geometric efficiency. This is attributed to the fact that 
the energy injection is effective for large radii, for which the structure of the emission
region is not relevant. The differences of the other parameter values, $p, \gamma_M,
r_2,$ and $\tan\theta_1$, affect the efficiency according to the simple formula Eq.~(\ref{eq:formula}).

In fact, $\gamma_M$ is typically much larger than $10^2$ (see Section~\ref{subsec:spectrum}), 
which may significantly reduce the efficiency. To check this, we plot the factor 
$\int^{r_2}_{\tilde{r}_1} \gamma^{-p+1} d\gamma/\int^{\gamma_M}_1 \gamma^{-p+1} d\gamma$ 
in Eq.~(\ref{eq:formula}) for $\tilde{r}_1 = 3$, $r_2=10$, and various $\gamma_M$ in 
Figure~\ref{fig:eff}. It is found that this factor for $\gamma_M > 10^4$ still can have a high value 
$\sim 0.25$ when the power-law index is $p \sim 2.3-3$. For $\gamma_M = 10^6$ and
$p=2.5$, the result of the numerical integration is 
\begin{equation}
\frac{L_j}{\dot{M}c^2} = 0.059\;f_n f_a f_{\rm th},
\label{eq:m8e}
\end{equation}
while the approximate estimate gives us $L_j/(\dot{M} c^2 f_n f_a f_{\rm th}) \sim 0.04$.
Note that this corresponds to the maximum level of the efficiency in this system, which is obtained
for the parameters $p=2.3-3, \tan\theta_1 = 1,$ and $r_2 = 10$. However, we consider that
such parameter values are within realistic ranges.

The numerical calculation gives us the mass injection efficiency for $\gamma_M = 10^6$ and $p=2.5$
\begin{equation}
\frac{\dot{M}_j}{\dot{M}} = 0.019\;f_n f_a f_{\rm th}.
\label{eq:m8m}
\end{equation}
This indicates that the injected protons will be isotropized and have mean random Lorentz factor 
$\langle \gamma \rangle = L_j/(\dot{M}_j c^2) = 3.1$. If no additional energy is injected into the polar 
region, the fireball model predicts that the random kinetic energy is transferred to the bulk kinetic 
energy with terminal Lorentz factor $\Gamma \simeq 3.1$.

\subsection{Dependence on BH mass}

We also perform calculations for various values of BH mass. For the case of $M_8=10$, 
the neutron emission region is set between $\rho_m = 10$ and $\rho_M=30$. 
We find that many of the properties for this case are just a scale-up version of those 
for $M_8=1$ discussed above. The energy is injected mainly for $\gamma \sim 30-100$
in the case of $r_2 = 100$. The energy injection efficiency is approximately
estimated by Eq.~(\ref{eq:formula}). For $p>2$, the energy in the range $\gamma \sim
30-100$ is smaller than that in $\gamma \sim 3-10$, so that we have smaller energy
injection in this case. Since the mean Lorentz factor of the injected protons 
$\langle \gamma \rangle$ is larger, the mass injection efficiency is even smaller.

\begin{figure}
\epsscale{1.15}
\plotone{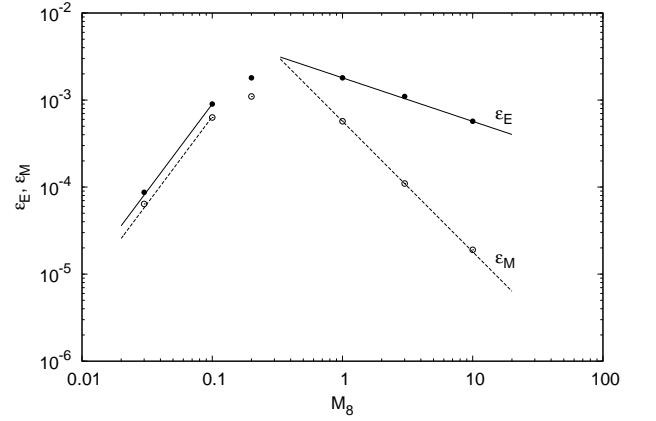}
\caption{
Results of the calculations of the energy injection efficiency $\epsilon_E = L_j/\dot{M}c^2$
(filled circles) and of the mass injection efficiency $\epsilon_M = \dot{M}_j/\dot{M}$ (open circles) 
as a function of $M_8 = M/(10^8\;M_{\odot})$ with the parameters 
$f_n f_a f_{\rm th} = 3\times 10^{-2}$, $\tan\theta_0=2$, $p=2.5$, $\gamma_M=10^6$,
$r_2 = 10R_{\rm s}$, and $\tan\theta_1=1$. For $M_8 \gtrsim 0.3$, $\epsilon_E \propto
M_8^{-p+2}$ (represented by solid line) and $\epsilon_M \propto M_8^{-p+1}$ (dashed line),
consistent with the approximate formula Eq.~(\ref{eq:formula}), while for $M_8 \leq 0.1$, 
$\epsilon_E$ and $\epsilon_M$ both $\propto M_8^2$ (solid and dashed lines), consistent
with the approximate formula Eq.~(\ref{eq:formula2}).
The final Lorentz factor of the jet (without an additional energy input) is given by 
$\Gamma = \epsilon_E/\epsilon_M$.
}
\label{fig:calc}
\end{figure}

We plot the calculated $L_j/(\dot{M}c^2)$ and $\dot{M}_j/\dot{M}$ for $M_8 = 3$ and $10$
with the same parameters as for Eqs.~(\ref{eq:m8e}) and (\ref{eq:m8m}) and 
$f_n f_a f_{\rm th} = 3 \times 10^{-2}$ in Figure~\ref{fig:calc}.
The approximate formula Eq.~(\ref{eq:formula}) implies that $L_j/\dot{M}c^2 \propto r_2^{-p+2}
\propto M_8^{-p+2}$ for given $\theta_1$ and $\gamma_M$, which agrees with the numerical
results. The mean Lorentz factor should linearly depend on the length scale, i.e., 
$\langle \gamma \rangle \propto M_8$, leading to $\dot{M}_j/\dot{M} = 
L_j/(\langle \gamma \rangle \dot{M} c^2) \propto M_8^{-p+1}$, which also agrees with the numerical
results. These scalings are applicable for cases of $\tilde{r}_1 > 1$, i.e., $M_8 \gtrsim 0.3$.

For the case of $M_8 = 0.1$, the system size is smaller than the case of $M_8=1$,
i.e., $\rho_m = 0.1$, $\rho_M=0.3$, and $r_2 = 1$. In this case most of the energy is 
injected into a region at $r_j > r_2$, which is assumed to be significantly collimated, so that 
the injection efficiency should be smaller than the case of $M_8 = 1$. Let us assume that the dilute
polar region is cylindrical at $r_j > r_2$ for simplicity, and then integrate $\dot{u}_\gamma^{(p)}$ 
and $\dot{u}_\gamma^{(p)}/(\gamma c^2)$ for the energy and mass injection, respectively, 
over $1 < \gamma < \gamma_M$ and the volume of the polar region at $r_j \geq 1$ to 
deduce the efficiencies. The results (and those for the cases of $M_8 = 0.03$ and $0.2$) 
are plotted in Figure~\ref{fig:calc}. 

We can also derive an approximate formula of the energy injection efficiency for $M_8 \leq 0.1$. 
In the cylindrical region, we have the opening angle as a function of $r_j$ as 
$\theta_c \simeq \theta_1 r_2/r_j$. The geometric efficiency may be estimated as 
\begin{eqnarray}
\frac{L_j}{f_n f_a f_{\rm th}\dot{M}c^2} &\sim& \frac{1}
{2 V \int^{\gamma_M}_1 \gamma^{-p+1} d\gamma} \nonumber \\
&& \times \int_1^{\infty} d\gamma \int^{\infty}_1 dr_j\; r_j^2 \left(\frac{\theta_c^2}{2}\right) 
\int d\Omega_j \gamma^{-p+1} \delta (r_j - \gamma) h \nonumber \\
&\sim& \frac{\theta_1^2 r_2^2}{4} \frac{\int^{\infty}_1 \gamma^{-p-1} d\gamma}
{\int^{\gamma_M}_1 \gamma^{-p+1} d\gamma} \simeq 0.03\;r_2^2, ~~~(r_2 \leq 1)
\label{eq:formula2}
\end{eqnarray}
where the final results are obtained for $p=2.5, \gamma_M=10^6$, and $\tan\theta_1=1$.
For $p\sim 2.3-3$ and $\gamma_M > 10^4$, the efficiency ranges between $0.02 r_2^2$
and $0.05 r_2^2$. This equation clearly indicates that the energy injection is smaller 
for smaller $M_8 (= 0.1\;r_2)$ in the case of $r_2 \leq 1$. The mean Lorentz factor may be 
estimated as 
$\langle \gamma \rangle = \int^{\infty}_1 \gamma^{-p-1} d\gamma/\int^{\infty}_1 
\gamma^{-p-2} d\gamma \approx (p+1)/p$. Those approximate formulae agree with the
numerical results of $L_j/\dot{M}c^2$ and $\dot{M}_j/\dot{M}$.

\section{Conclusion and Discussion}
\label{sec:discussion}

The luminosity of the relativistic neutrons from geometrically thick AGN accretion flow is 
estimated as $L_n = f_n f_a f_{\rm th} \dot{M} c^2$, and $f_n f_a f_{\rm th}$ can be as high as 
$\sim 3 \times 10^{-2}$, where the thermalization fraction of the accretion power 
$f_{\rm th} \lesssim 0.3$, the energy fraction of the accelerated protons in the thermal energy 
$f_a \lesssim 0.3$, and the neutronization (and escape) fraction $f_n \lesssim 0.3$.
Those neutrons escape isotropically, a fraction of which decay into protons at the 
dilute polar region and inject the energy and mass into the relativistic jet.
We have calculated this geometric fraction by setting a simple system consisting of 
the central BH, the accretion flow including the neutron emission region 
(at $R_{\rm s} \leq \rho \leq 3 R_{\rm s}$), and the polar region. The polar region is assumed
to be a cone with an opening angle of $\theta_1 \sim {\it O}(1)$ at $r_1 = R_{\rm s} < r_j < r_2 =
10 R_{\rm s}$ and changed to be a cylinder, at $r_j > r_2$ as a rough approximation of the 
significant collimation.

The results are plotted in Figure~\ref{fig:calc}, for which the parameters
are chosen as $\tan\theta_0=2$, $r_2 = 10R_{\rm s}$, $\tan\theta_1=1$, 
$p=2.5$, $\gamma_M=10^6$, and $f_n f_a f_{\rm th} = 3\times 10^{-2}$, and 
they are well approximated by Eqs.~(\ref{eq:formula}) for $M_8 \gtrsim 0.3$ and (\ref{eq:formula2})
for $M_8 \leq 0.1$. 
The efficiencies $L_j/\dot{M}c^2$ and $M_j/\dot{M}$ do not significantly depend on the torus 
thickness characterized by $\tan\theta_0$, because the energy injection is effective for large 
radii, for which the structure of the emission region is not relevant. The efficiencies are larger
for the larger polar region, i.e., larger $r_2$ and/or $\theta_1$, according to Eqs.~(\ref{eq:formula}) 
and (\ref{eq:formula2}). They are also weak functions of 
$p$ and $\gamma_M$ for $2.3 < p < 3$ and $\gamma_M > 10^4$, which are considered to be
realistic ranges.

For $M \sim 10^{8}\;M_{\odot}$, this process can produce a relativistic jet with
$L_j \sim 2 \times 10^{-3}\;\dot{M} c^2$ and $\dot{M}_j \sim 6 \times 10^{-4}\;\dot{M}$,
leading to the final Lorentz factor $\Gamma \sim 3$ (if no other types of energies are injected). 
Most of the relativistic neutron luminosity $L_n \sim 3\times 10^{-2}\;\dot{M} c^2$ 
is injected outside the polar region, which may contribute to the dense non-relativistic disk 
wind energy. The existence of such energetic disk winds is not incompatible with observations 
\citep{tombesi11,tombesi12}. The main part of the accretion power $\dot{M} c^2$ is carried by
protons and magnetic fields, which may also contribute to the relativistic jet and the non-relativistic
disk wind, or just be swallowed by the central BH.

The observations suggest that the luminosities of AGN jets are broadly distributed
over $L_{j,{\rm tot}} \lesssim \dot{M} c^2$ \citep[e.g.,][]{fernandes11,punsly11,ghisellini10}. 
Our results imply that jets with $L_{j,{\rm tot}} \lesssim 2 \times 10^{-3}\;\dot{M} c^2$ may 
be produced by the neutron decay process by itself. In this model, larger BH masses are 
associated with larger final Lorentz factors of the jets for $M > 10^8\;M_{\odot}$.

For AGN jets with higher luminosities, the neutron decays still can be a dominant mass loading
process, whereas they can have additional energy inputs, such as Poynting flux. The mass loading
$\dot{M}_j \sim 6 \times 10^{-4}\;\dot{M}$ leads to e.g., $\Gamma \sim 50$ for
$L_{j,{\rm tot}} \sim 3 \times 10^{-2}\dot{M}c^2$.
If the additional energy input scales as $L_{j,{\rm tot}} \propto M$, jets with larger $\Gamma$ are 
associated with smaller $M$ for $M < 10^8\;M_{\odot}$ and with larger $M$ for $M > 10^8\;M_{\odot}$.

We have simplified the configuration of the BH accretion system, approximated the 
amount of the created protons as a delta function of the traversing distance of neutrons
(Eq.~\ref{eq:delta}), and assumed the Euclidean space, to obtain orders of magnitudes of the 
geometric efficiencies. This study has provided the useful approximate formulae Eqs.~(\ref{eq:formula})
and (\ref{eq:formula2}).
More sophisticated formulations and calculations, taking into account the collimating shape
of the dilute polar region, radial dependence of the neutron emissivity, detailed process 
of the neutron decay, and geometry around the Kerr BH, are worth investigating in separate papers.

While we have focused on the geometrically thick disks in this paper, the proton acceleration
and the relativistic neutron production may be effective also in hot coronae above the 
geometrically thin disks \citep[e.g.,][]{drury12,degouveia10,vieyro12}. Similar scalings of the 
energy and mass injections to the outflows
are expected. In any cases, it is important to discuss the mechanism for regulating 
$\Gamma \sim 10-100$ of AGN jets and the role of the neutron component.

\acknowledgements
We thank the referee for useful comments.
K.T. thanks M.~Sikora and A.~Janiuk for useful discussions and Nicolaus Copernicus 
Astronomical Center for the hospitality during his stay. This work is partly supported by 
JSPS Research Fellowships for Young Scientists No. 231446 (K.T.) and by KAKENHI
20540231 (F.T.).

\end{document}